\baselineskip=7truemm

\def\v{\vskip 2mm}

\def\noi{\noindent}
\def\no{\noindent}
\def\tfr{{Tully-Fisher relation}}
\def\kms{km s$^{-1}$}
\def\endpage{\vfil\break}
\def\Deg{^\circ}

\def\r{\hangindent=1pc  \noindent}
\def\cen{\centerline}

\def\co{$^{12}$CO($J=1-0$)}
\def\ta{$T^*_{\rm A}$}
\def\micron{$\mu$m}

\topskip 30mm

\cen{\bf DISTANCE MEASUREMENT OF GALAXIES TO REDSHIFT OF  $\sim$ 0.1 USING}
\cen{\bf THE CO-LINE TULLY-FISHER RELATION}

\vskip 20mm

\cen{Yoshiaki SOFUE$^{1,2}$, Franz  SCH{\"O}NIGER$^1$, Mareki HONMA$^1$, 
Yoshinori TUTUI$^1$,}

\cen{Takashi ICHIKAWA$^2$, Ken-ichi WAKAMATSU$^3$,}

\cen{Ilya KAZES$^4$, and  John DICKEY$^5$}

\vskip 10mm

\cen{\it E-mail: sofue@mtk.ioa.s.u-tokyo.ac.jp}

\vskip 10mm

\cen{\it (1) Institute of Astronomy, University of Tokyo, Mitaka, Tokyo 181, Japan}

\cen{\it (2) Kiso Observatory, IoA, University of Tokyo, Kiso-gun, Nagano 397-01, Japan}

\cen{\it (3) Dept. Physics, Gifu University, Gifu, 501-11, Japan}

\cen{\it (4) Observatoire de Paris, F-92190 Meudon, France}

\cen{\it (5) Dept. Astronomy, Univ. Minnesota, 116 Church St. SE,}
\cen{\it  Minneapolis, MN 55455,  USA}

\vskip 20mm
\cen{(to appear in PASJ, 1996)}

\topskip 0mm
\endpage

\centerline{\bf Abstract:}

We report on the first results of a long-term project to derive distances 
of galaxies at cosmological distances by applying the CO-line
width-luminosity relation.
We have obtained deep CO-line observations of galaxies at redshifts up to
$cz \sim 29,000$ \kms\ using the Nobeyama 45-m mm-wave Telescope, and
some supplementary data were obtained by using the IRAM 30-m telescope.
We have detected the CO line emission for several galaxies, and used their
CO line widths to estimate the absolute luminosities using the 
line-width-luminosity relation.
In order to obtain photometric data and inclination correction, we also 
performed optical imaging observations of the CO-detected galaxies
using the CFHT 3.6-m telescope at high resolution.
The radio and optical data have been combined to derive the distance
moduli and distances of the galaxies,  and Hubble ratios
were estimated for these galaxies.
We propose that the CO line width-luminosity relation can be a
powerful method to derive distances of galaxies to redfhift of $z\sim 0.1$
and to derive the Hubble ratio in a significant volume of the universe.

\v\v
\noi{\bf Key words:} Cosmology --  Galaxies: general --
Distance scale  -- Radio lines: CO molecular

\v\v\v

\noi{\bf 1. Introduction}\v

The HI-line width - luminosity relation (\tfr)
is one of the most powerful tools to measure 
distances to galaxies (Tully \& Fisher 1977; Aaronson et al. 1986;  Pierce 
\& Tully 1988; Kraan-Korteweg et al. 1988;  Fouqu{\'e} et al. 1990; 
Fukugita et al. 1991).
However, distances to galaxies so far reached by HI observations are limited
to around 100 Mpc, or 
$cz \sim 10,000$ to 15,000 \kms, even
with the use of the world's-largest telescopes (Sch\"oniger and Sofue 1993).
We have no routine method to determine distances to galaxies beyond
this distance, at which  beam sizes of a few arc-minutes
at $\lambda$ 21 cm  become too large to resolve individual 
galaxies in a cluster. 
Interferometers like the VLA are not very useful for this purpose 
because of the limited number of spectral channels (velocity resolution).
Furthermore,  the red-shifted HI frequency results in
increases in beam size as well as increased  interference, 
which also makes observations of distant cluster galaxies difficult.
Moreover, HI line profiles are easily disturbed by interactions
among galaxies, which is inevitable in the central region of a cluster,
causing uncertainty in the interpretation of  HI line profiles for the \tfr.
On the other hand, molecular gas is tightly confined to the luminous 
stellar disk, so that it is less affected by tidal interactions and by 
ram-pressure distortion due to the intra-cluster gas.
The molecular gas is distributed to a radius
of several to ten kpc, so that the integrated line profiles manifest the 
maximum velocity part of the rotation curve (Sofue 1992).

Molecular-line observations at millimeter wavelengths, particularly in the
CO-line emission at 115 GHz, can be achieved with  much sharper beams.
Therefore,  we will be able to resolve individual member galaxies
in a cluster more easily, which makes it possible to avoid 
contamination by other member galaxies in a beam.
Moreover, the larger the redshift of an object, the lower the
CO frequency, which results in a decrease in the system noise temperature
due to the atmospheric O$_2$ emission at 118 GHz:
the more distant is a galaxy, the lower is the noise temperature.

Dickey and Kazes (1992) have addressed the use of CO line widths
instead of and/or supplemental to HI observations, and proposed the
CO-line Tully-Fisher relation as an alternative to the HI \tfr.
The CO-line width - luminosity relation has been established for the local 
distance calibrators and tens of nearby galaxies (Sofue 1992;  Sch{\"o}niger 
and Sofue 1993, 1996).
In these works, we have shown that CO-line measurements can be 
used as  an alternative  to HI by deriving a good linear correlation between 
CO and HI linewidths for the galaxies in the sample.
The only disadvantage of the use of CO line is the sensitivity. 
Actually, we need a few mK rms noise for line-width measurements
at a velocity resolution of 10 \kms\ for normal galaxies beyond 
$cz \sim 10,000$ \kms, which requires long integration times.
Such observations are possible only by a long-term project
with the use of the world's-largest mm-wave telescopes.

On the basis of these studies, we have conducted a long-term project to
observe  $^{12}{\rm CO}(J=1-0)$ line profiles for distant cluster galaxies
at redshift to $z \sim 0.1$ using the Nobeyama 45-m telescope.
The targets are so chosen that they include galaxies in clusters
of galaxies at $cz \sim 10,000$ to 30,000 \kms.
Before proceeding to measuring CO in normal galaxies at these
distances, we chose far-infrared (FIR) luminous galaxies, 
so that the CO detection is easier at this first trial.
We selected the targets using the NED (Nasa Extragalactic
Database) searching facility.
In this paper, we present results for the most FIR luminous
galaxies with 60 \micron\ and 100 \micron\ fluxes greater than
0.2 and 0.5 Jy, respectively,
as well as with  redshifts within 10,000 and 30,000 \kms.
This sample selection may include some danger in applying the
Tully-Fisher relation because of possible peculiarities in optical
absolute magnitudes for the FIR luminous galaxies:
We may need to evaluate the accuracy of the relation for such
FIR luminous galaxies, which remains problem for the future.
The result presented here should be taken as a demonstration of
the possible usage of the CO Tully-Fisher relation for measuring distances of
galaxies of these large distances.   
However, we mention that such a FIR luminous starburst galaxy
NGC 253 is included in the small number of standard calibrators of
the Tully-Fisher relation.
It is also known that IRAS galaxies obey a standard 
Tully-Fisher relation (van Driel et al 1995).

We obtained additional CO-line spectra for some galaxies using the IRAM 30-m
mm-wave telescope.
We also obtained high-quality optical imaging of detected galaxies in CO
in order to measure the inclination and magnitude using the 
Canada-France-Hawaii 3.6-m telescope.
In this paper,  we report a first-step result of the measurement
of CO line profiles in order to demonstrate that the line profiles can
be obtained in  routine work within a reasonable integration time,
and that distances can be obtained by applying the \tfr. 
We also report on optical imaging observations of several CO-detected
galaxies, and  try to obtain a possible value of the Hubble ratios.

\v\v\v
\no{\bf 2. CO-Line Observations}\v

\v\no{\it 2.1. Observations}\v

We have selected spiral galaxies which satisfy the following criteria using 
NASA Extragalactic Database (NED); 
1) galaxies with red-shift between 10,000 $\rm km\; s^{-1}\;$ 
and 30,000 $\rm km\; s^{-1}\;$,
2) with position error less than 10$''$,
3) with relatively strong far infrared emission at 60 $\mu$m and 100 $\mu$m.
The first criterion was set to select distant galaxies for which the 
HI Tully-Fisher relation cannot reach.
The highest redshift galaxy was IRAS 08344+5105 which had
$cz=29029$ \kms.
The second criterion was set so that the position error is small 
enough compared to the beam size of CO-line observation.
Accurate positions of the sources were measured on the Palomar Sky Survey
Prints using the position measuring facilities at the Kiso Observatory,
as well as using the STScI Digitized Sky Survey images.  
The third criterion was to select galaxies 
which are expected to be bright in the CO-line emission.
One might suspect that galaxies with bright infrared emission 
could deviate from the Tully-Fisher relation for normal galaxies.
However, van Driel et al.(1995) studied so-called IRAS Minisurvey 
galaxies and showed that there is no significant difference between 
Tully-Fisher relation for IRAS selected galaxies and that for 
normal galaxies.
Among galaxies which satisfy these criteria, 
interacting galaxies and Seyfert-type galaxies are excluded.
Galaxies with larger error of recession velocity than 
100 $\rm km\; s^{-1}\;$ are also excluded, though galaxies for 
which the error of recession velocity is unknown are included.
There are some galaxies for which types are not known through NED.
For such galaxies, we have confirmed that they are disk galaxies 
and do not show interaction on the Palomar Sky Survey Prints.

Observations of the \co\ line of distant cluster galaxies 
were made on 1994 January 14 to 23, 1994 December 9 to 12,
and 1995 January 6 to 10,  March 13 to 17,  
1995 December 17, 18, 21, 22, and 1996 February 18 to 22, using 
the 45-m telescope of the Nobeyama Radio Observatory.
The antenna had a HPBW of $15''$ at the CO line frequency, 
and the aperture and  main-beam 
efficiencies were  $\eta_{\rm a}=0.35$ and $\eta_{\rm mb}=0.50$, respectively.
We used two  SIS (superconductor-insulator-superconductor)
receivers with orthogonal polarization
which were combined with 2048-channel acousto-optical spectrometers.
The total channel number corresponds to  a frequency width of 250 MHz,
and, therefore, to a velocity coverage in the rest frame at the galaxy 
of $c~ 250 {\rm MHz}/\nu_{\rm obs}
=c(1+z)$(250 MHz/115.2712 GHz)=650$(1+z)$ \kms.
The center frequency was so tuned that the center channel 1024
corresponds to 115.2712$(1+z)$ GHz for each galaxy.

After combining every 32 channels in order to increase the signal-to-noise 
ratio, we obtained spectra with a velocity resolution of 10.2 \kms. 
The system noise temperature (SSB) was 300 to 400 K at the observing frequencies.
The calibration of the line intensity was made using an absorbing chopper 
in front of the receiver, yielding an antenna temperature (\ta), corrected 
for both the atmospheric and antenna ohmic losses.
We used an on-off switching mode, and the 
on-source total integration time was 2 to 6 hours for each galaxy.
After careful flagging and subtraction of linear baselines, 
the rms noise of the resultant spectra 
at velocity resolution of 10 \kms\ was typically 2 mK in \ta.
The pointing of the antenna was tested by observing nearby SiO maser sources 
at 43 GHz every 1 to 1.5 hours, and was typically within $\pm 3''$ during  good
weather conditions which were attained for about one fourth of the allocated
observing time.

\v\no{\it 2.2. CO-line Profiles and Results}\v

Among the observed galaxies, sufficient quality data have been obtained for
fifteen galaxies including eight galaxies with CO-line detection.
Good CO line profiles were obtained for several galaxies with a sufficient 
signal-to-noise ratio for determining the velocity width.
CO-detected galaxies are 
NGC 6007, IZw 23, IC2846, CGCG 1448.9+1654, CGCG 1113.7+2936,
CPG 60451, CGCG 1417.2+4759, and IRAS 17527+6422.
We obtained possible (marginal) detection for two galaxies,
IRAS 14210+4829 and IRAS 23420+2227.
These galaxies are listed in Table 1, and the spectra are shown in Fig. 1. 
The intensity scale used in this paper is the antenna
temperature ($T_{\rm A}^*$).
The abscissa in Fig. 1 is relative recession velocity $V_0$
with respect to the optical redshift ($cz$) given in table 1.
We have also performed CO line observations using the IRAM 30-m
Telescope, and  obtained  detection for CGCG 1417.2+4759.
The observation with the IRAM 30-m telescope was made on July 9, 1994,
equipped with an SIS receiver of system temperature of 270 K and
combined with a 256-channel filter-bank
spectrometer of a velocity coverage 1400 \kms.
The CO profile from the IRAM 30-m telescope is also shown in Fig. 1.

The profiles for the detected galaxies show a nearly double-peaked
emission, which is characteristic of a rotating disk.
From these data we have measured the line width,
peak antenna temperature, and line intensity for individual galaxies.
The results are listed in Table 2.
The farthest galaxies for which the CO-line has been detected 
in the present program are IRAS 23420+2227 and IRAS 17527+6422 at 
$cz=26022$ and 26151 \kms, respectively.

\v
\cen{--Fig. 1;  Tables 1, 2--}
\v

Our program included  farther galaxies with redshift to $cz=29000$ \kms,
and good data were obtained for several galaxies after integration of a few hours.
Some examples are shown in Fig. 2, which are for 
IRAS 16305+4823 with $cz=26327$ \kms, showing a slight enhancement at
relative velocity of $V_0 \sim -180$ to 0 \kms.
IRAS 07243+1215 with $cz=28204$ \kms, showing marginal detection at
$V_0 \sim +70$ to +200 \kms, while not conclusive; 
IRAS 08344+5105 with $cz=29029$ \kms, the farthest galaxy in our program
showing a slight sign  of enhancement at $V_0 \sim -100$ to +180 \kms.

\v
\cen{--Fig. 2--}

\v\v\v
\no{\bf 3. Optical Observations}\v

\v\no{\it 3.1. Observations}\v

In order to obtain photometry, morphology and inclination
of the CO detected galaxies, we have performed
high-quality optical imaging observations of three galaxies,
CGCG 1417.2+4759, CGCG 1448.9+1654, and CPG 60451.
The galaxies were so chosen that they are among galaxies showing the
best quality CO line profiles observed during the 1994/1995 missions,
and are within a right ascension range observable during
the allocated time of optical observation.
The observations were made using the
Canada-France-Hawaii 3.6-m Telescope (CFHT) on June 30 - July 1, 1995.
Additional images have been taken with the 105-cm Schmidt telescope
of the Kiso Observatory in January 1994.
Using these data, we have determined the apparent magnitudes
and inclinations of the galaxies.
We summarize the optical data in Table 3.

\v
\cen{--- Table 3 ---}
\v

Imaging observations were made in Johnson $V$ and Kron-Cousins 
$R_{\rm C}$ and  $I_{\rm C}$ bands at the Canada-France-Hawaii 3.6-m
Telescope with the SIS (Subarcsecond-Imaging-Spectrograph).
The Loral3 CCD had an effective imaging area 
$2048 \times 2048$ pixels with a pixel size of 
$15\mu {\rm m} \times 15 \mu{\rm m}$.
The camera covered a sky area of $3' \times 3'$ with a resolution of 
$0.086''$ per pixel. 
The exposure times in the $V$, $R_{\rm C}$, and $I_{\rm C}$
bands were 900, 900, and 300 sec, respectively.
The observations were performed under photometric sky conditions on 30 June,
1995. 
Standard stars from Landolt (1992) were also observed for flux 
calibration. 
The FWHM seeing size was $0.6 - 0.8''$, which is small enough to 
determine the morphology of the observed galaxies.
The dome screen was exposed to obtain flat frames.

Standard data processing (bias subtraction and flat fielding) was performed 
with the IRAF software package. 
For further image processing and surface photometry (e.g., sky subtraction 
and flux calibration) we used the SPIRAL package developed at Kiso Observatory 
and installed into the IRAF system. 
The large field of view gives a sufficiently large sky area around the 
galaxies to allow an accurate sky-subtraction.
The flux calibration using the standard star frames resulted in a
photometric accuracy of about $\pm3$\%.
The transformation to Johnson and Kron-Cousins' standard photometric
systems and the atmospheric extinction corrections were determined by 
analyzing the standard star frames.

\v\no{\it 3.2 Surface Photometry}\v

In Fig. 3 we present $R_{\rm C}$ band images of three galaxies.
The contours are drawn at an interval of
1 mag./square arcsec, starting from 26.5 mag./square arcsec.
The  maps were fitted with ellipses to obtain the 
isophotal magnitudes and the axial diameters.
The second column of Table  3 gives morphological types of the galaxies.
We stress that the galaxies show quite normal morphology as spirals,
and are isolated galaxies.
The isophotal magnitudes of the galaxies above 25.5, 26.0,
24.5 mag arcsec$^{-2}$ levels in the $V$, $R_{\rm C}$, and $I_{\rm C}$
bands are presented, respectively.
The major $A$ and minor $B$ axial diameters at 25.5 mag arcsec$^{-2}$ in
the $V$ band are given in the last two columns.
The typical error in the total magnitude estimate was about 0.03 magnitude.
 
\v\cen{ -- Fig. 3 --}\v
 
We measured the intensity of the galaxies in elliptic annuli of one pixel
width centered on the galactic nucleus, after subtracting the sky brightness
and stellar images, thus obtaining a growth curve.
Since the integrated magnitude in the tail of each growth curve
approximately follows an exponential law, asymptotically reaching
a constant value, we define here the total magnitudes as
the extrapolation of this curve to infinity.

To evaluate the@\tfr\ we need the velocity width to be corrected
for the inclination $i$ and for redshift effect to an edge-on value at
the rest  frame.
We obtain an intrinsic velocity width in the rest frame referred to the
galaxy corrected for the inclination by
$$  W_{i} = W/{\rm sin}~ i, \eqno(1)$$
where $W$ is the rest-frame velocity width given by
$$ W=W_{\rm rest}= c \Delta \nu_{\rm obs}/\nu_{\rm obs}
 = c (1+z) \Delta \nu_{\rm obs}/(115.2712~{\rm GHz}), \eqno(2)$$
where $ \Delta \nu_{\rm obs}$ is the observed line width in frequency.
We obtain the inclination $i$ of a galaxy from the conventional formula given
by Hubble (1926) for oblate spheroid, 
$${\rm cos}^2i= (q^2-q_0^2) / (1-q_0^2) \eqno(3)$$
with $q=b/a$ and $q_0=c/a$, where $a$, $b$, and $c$ are the lengths of three 
axes of the spheroid.
Here, we adopt $q_0  = 0.20 $ for the present analysis.
The measured result is given in Table 4.
The typical error in the estimate of inclination was $\pm 3\Deg$.

The Galactic absorption $A_{\rm B}$ is taken from  Burstein \& Heiles (1984).
The internal absorption correction $A_i$ are computed by the methods given in 
Pierce \& Tully (1992).
For the K-correction, we take the correction of Fukugita, et al (1995).
The corrected total magnitude is written as
$$m_{\rm T}^{b,i} = m_{\rm T} - A_B - A_i - K.\eqno(4)$$
Obtained results  of the optical measurements are summarized in Table 4.
Typical errors in the magnitude estimate was about $\pm 0.1$ magnitude
except for the systematic errors which might be present in the applied 
corrections.

\v\cen{-- Table  4 --}\v

\no{\bf 4. Discussion}

\v\no{\bf 4.1. Distances and Hubble Ratios}\v

On the basis of our previous study that the CO-line width of a galaxy 
is equivalent to the HI-line width within the error of the \tfr\
for local calibrators, we try to derive the distances to the three
galaxies for which the optical photometry and imaging have been obtained
with the CFHT.
We adopt the zero point of the TF relation given by Pierce \& Tully (1992)
for $R$ and $I$ and by Shimasaku \& Okamura (1992) for $V$ band.
   $$M_{V}^{b,i} = -20.43 - 6.21 ({\rm log} W_{i} - 2.5)\eqno(5) $$ 
   $$M_{R}^{b,i} = -20.40 - 8.23 ({\rm log} W_{i} - 2.5) \eqno(6)$$ 
   $$M_{I}^{b,i} = -20.94 - 8.72 ({\rm log} W_{i} - 2.5) \eqno(7)$$ 
We adopt the same scheme for their extinction and inclination corrections.
However, the distance moduli of  the common local calibrators exhibit large 
discrepancies. 
The differences are $-1.0$ to +0.1 mag and 0.38 mag on average.
(Shimasaku \& Okamura's moduli are larger).
Four calibrators (M31, M33, NGC300, and NGC2403) 
have  distances determined well by Cepheid variables.
The Cepheid distances are in favor of Shimasaku \& Okamura's values.
Therefore, we adopt the Shimasaku \& Okamura's calibration and add 
19\%to the distance moduli obtained by the Pierce \& Tully 
calibration  in $R$ and $I_{\rm C}$ bands.

Using the \tfr\ for the CO-line widths measured in section 2
we derive  distance moduli, distances and Hubble ratios 
$V_{\rm C}/r$ for the three galaxies.
Here the recession velocity $V_{\rm C}$ is the value with respect to the
rest frame referred to the cosmic background radiation (Smoot et al. 1991).
Distance modulus, distance, and Hubble ratio obtained in individual 
optical  bands ($V$, $R_{\rm C}$ and $I_{\rm C}$) are shown in Table 5 for each object.
Error in the absolute magnitude $\Delta M$ 
arises from the error in CO line-width measurement $\Delta W$ as well
as from that due to the error in inclination $\Delta W_i$, and is given by 
$$ \Delta M \simeq \simeq
K \sqrt{(\Delta W/W)^2 + (\Delta W_i/W_i)^2}, \eqno(8)$$
where $K$ is the coefficient of the \tfr\ in equations (5) to (7), and
$$\Delta W_i=W {{\rm cot}~i \over {\rm sin}~i }\Delta i 
= W_i {\rm cot}~ i \Delta i. \eqno(9)$$
Here, $\Delta i$ is the error in inclination measurement, which was typically
$\pm 3\Deg$.
The distance error is then given by  
$$ {\Delta D \over D} \simeq {\Delta (m-M) \over 5 ~{\rm log}e} 
= {1 \over 5~ {\rm log}e} \sqrt{\Delta m^2 + \Delta M^2}. \eqno(10)$$
Here $\Delta m$ is the error in the photometric measurement
of apparent magnitude.
In the present case the error in the apparent magnitude ($\sim 0.1$ mag.) 
is much smaller than that of the estimate of absolute magnitude 
arising from the line-width measurement and inclination correction.
Hence, we have
$$ {\Delta D \over D} \simeq {\Delta M \over 5~ {\rm log}e}\simeq {K \over 2.17}\Delta W_i/W_i 
\simeq {K \over 2.17}\sqrt{ (\Delta W/ W)^2 + ({\rm cot}~ i ~ \Delta i)^2 }
\eqno(11)$$

The measured values in the $V$, $R_{\rm C}$, and $I_{\rm C}$ band  are shown in Tables 4 and 5.
The errors shown here do not include systematic error in the adopted photometric
corrections or the internal scatter within the \tfr\ itself.
As to the CO line width of CGCG 1417.2+4759 we adopted the CO data from IRAM 30-m telescope.
The Hubble ratio obtained for CGCG 1417.2+4759 from the NRO 45-m telescope
data is as large as 110 \kms/Mpc, while the data from IRAM 30-m
gives a value around 80, and we here adopt the value from IRAM data.
The larger value of Hubble ratio from the NRO data appears to be due to
an underestimate of the velocity width because of the smaller
velocity coverage in the spectrometer at NRO than the IRAM 30-m telescope.
The obtained Hubble ratios in table 5 are scattered
from $\sim 50$ to $\sim 80$ \kms/Mpc, while the values are within a
reasonable range of current determinations.

\v\cen{-- Table  5 --}\v

\v\no{\bf 4.2. Remarks}\v

We have underway a long-term program using the Nobeyama
45-m mm-wave telescope, and partly using the IRAM 30-m telescope,
to study the CO-line width-luminosity (CO Tully-Fisher) relation. 
In this  program, we have shown that the present method  
can be  a useful technique to derive distances of galaxies 
at redshifts greater than $cz \sim 10,000$  \kms,
at which the HI-line \tfr\ becomes more difficult to be applied.
Although the number of CO detection is not sufficient
for conclusive determination of the Hubble ratio, we have obtained
reasonable values for a few galaxies.
We indeed tried to observe more distant galaxies, whose redshifts 
were as large as $z \sim 0.1$, in January and February of 1996. 
We obtained marginal detection for a few galaxies as listed
in Table 1 and 2, and their CO line profiles are shown in Fig. 1 and 2.
In order to derive the Hubble ratios of these galaxies, however, 
high-resolution optical imaging as shown in Fig. 3 is required, 
which will be performed in the near future.

Our present sample selection may include some danger in applying the
Tully-Fisher relation because of possible peculiarities of FIR luminous
galaxies, although the optical imaging (Fig. 3) has shown that
the three galaxies used here are of normal morphology.
However, we mention that FIR luminous galaxies from the IRAS minisurvey
follow the Tully-Fisher relation (van Driel et al 1995).
We also note that such a FIR luminous starburst galaxy
as NGC 253 is included in the small number of standard calibrators of
the Tully-Fisher relation, indicating that the FIR luminosity does
not necessarily disturb this relation.

\v
{\bf Acknowledgment:} 
The observations were performed under a long-term
project at the Nobeyama Radio Observatory.
The authors would like to express their sincere thanks to 
Prof. N. Ukita and N. Nakai of the NRO for 
their kind and efficient help during the observations  
as well as for their invaluable suggestions. 

\v\v
\noi{\bf References} \v

\r Aaronson M., Bothun G., Mould J., Shommer
   R. A., Cornell, M. E., 1986, ApJ 302, 536

\r Aaronson, M., Mould, J., Huchra, J. 1980, ApJ, 237, 655

\r Burstein, D., Heiles, C. 1984, ApJS, 54, 33

\r Dickey J., Kazes I., 1992, ApJ 393, 530

\r Fukugita M., Okamura S., Tarusawa K., et al., 1991, ApJ 376, 8

\r Fukugita, H., Shimasaku, K., Ichikawa, T. 1995, PASP, 107, 945.

\r Fouqu{\'e} P., Bottinelli L., Gouguenheim L., 1990, ApJ 349, 1

\r Hubble, E. P. 1926, ApJ, 64, 321

\r Kraan-Korteweg R. C., Cameron L. M.,  Tammann G. A., 1988, ApJ 331, 610

\r Landolt, A. U. 1992, AJ, 104, 340

\r Pierce M. J., Tully R. B., 1988, ApJ 330, 579

\r Pierce, M. J., Tully, R. B. 1992, ApJ, 387, 47

\r Sch\"oniger, F., Sofue, Y. 1993, AA, 283, 21.

\r Sch\"oniger, F., Sofue, Y. 1996, AA, in press.

\r Shimasaku, K., Okamura, S. 1992, ApJ., 398, 441

\r Smoot, G. F., et al. 1991, ApJ,371, L1
  
\r Sofue Y., 1992, PASJ 44, L231

\r Tully B., Fisher J. R., 1977, AA 64, 661


\r van Driel, W., van den Broek, A, C., Baan, W. 1995 ApJ 444, 80.

\endpage

\settabs 15\columns
\+ Table 1. Galaxies detected in the CO line emission$^\dagger$.\cr
\vskip 8mm 
\hrule
\vskip 2mm

\+Name&&&&RA(1950)&&Dec(1950)&&&$cz$&&$S_{60\mu{\rm m}}$&$S_{100\mu{\rm m}}$
&Type&detection\cr
\+&&&&(h~ m~ s)&&($^\circ$~$'$ ~$''$)&&&(\kms)&&(Jy)&(Jy)&\cr
\+&&&&&&(posi. err. $"$)&&&(err. \kms) \cr
\vskip 2mm

\hrule
\vskip 2mm

\+ (1993/94 program)\cr
\vskip 2mm
\+CGCG 1113.7+2936&&&&11 13 46.9&&+29 35 59(4)&&&13880(33)&&0.63&2.03&?&D\cr
\+CGCG 1448.9+1654&&&&14 48 54.4&&+16 54 02(4)&&&13700(51)&&0.14&1.13&Sp&D \cr
\+CGCG 1417.2+4759&&&&14 17 14.8&&+47 59 00(2)&&&21465&&0.62&1.54&SBb&D \cr
\vskip 2mm
\+ (1994/95 program) \cr
\vskip 2mm
\+NGC 6007&&&&15 51 01.6&&+12 06 27(10)&&&10547(6)&&0.69&2.03&SBbc&D\cr
\+CPG 60451&&&&17 30 00.6&&+20 09 48.6(5)&&&14989(?)&&0.48&1.36&Scd&D\cr
\+IC 2846&&&&11 25 24.804&&+11 26 00.5(1)&&&12294(33)&&4.21&6.72&Sp&D \cr
\vskip 2mm
\+(1995/96 program) \cr
\+IRAS 17527+6422 &&&&17 51 44.99&&+64 22 14.1(3)&&&26151(24)&&2.22&3.25&?& D\cr
\+IZw 23 &&&& 09 56 01.0 && +52 29 48.0(1)&&&12224(57)&&0.62&1.74&Sp&D\cr
\vskip 2mm
\+ (Marginal detection)\cr
\vskip 2mm
\+IRAS 14210+4829&&&&14 21 06.2&&+48 29 59.0(1)&&&22690(?)&&0.38&0.88&Sp&?\cr
\+IRAS 23420+2227&&&&23 42 00.60&&+22 27 49.8(3)&&&26022(?)&&1.41&2.07&?&?\cr
\vskip 2mm
\hrule
\v
$\dagger$ Sources of the positions and redshifts is the NED (NASA Extragalactic
Database). These velocities were used for the center frequencies of
the CO-line detectors.

\vskip 20mm

\endpage

\settabs 11\columns
\+ Table 2. CO-line data for detected galaxies.\cr
\vskip 8mm 
\hrule
\vskip 2mm

\+Name&&& $cz$&  $cz_{\rm CO}$ && $\Delta \nu_{\rm obs}$ &$W~^\dagger$
&\ta&$I_{\rm CO,rest}$ &Type \cr
\+    &&& (\kms)&~~(\kms)  &&  (GHz)  &(\kms) & (mK)  & (K \kms)\cr
\vskip 2mm

\hrule
\vskip 2mm

\vskip 2mm
\+NGC 6007&&&   10547 &10542$\pm5$&& 0.129      & 347$\pm10$  & 23$\pm3$ & 5.1$\pm0.2$& SBbc\cr
\+IZw 23 &&&    12224 &12364$\pm5$&& 0.0391      & 106$\pm10$ & 37$\pm5$ & 2.4$\pm0.2$ & Sp \cr
\+IC 2846&&&    12294 &12302$\pm10$&& 0.0898      & 243$\pm15$  & 25$\pm4$   & 5.5$\pm0.2$&Sp\cr
\+CGCG 1448.9+1654&&& 13700&12693$\pm5$&& 0.0989 & 269$\pm10$ & 20$\pm3$& 3.0$\pm0.1$ & Sp&\cr
\+CGCG 1113.7+2936&&& 13880 &13910$\pm5$&&0.117& 318$\pm5$ & 16$\pm2$ & 2.5$\pm0.1$ & SBb \cr
\+CPG 60451&&&  14989 &14984$\pm5$&& 0.168    & 460$\pm15$ & 25$\pm3$  & 6.3$\pm0.2$ & Scd\cr
\+CGCG 1417.2+4759&&& 21465&21527$\pm10$&& 0.111   & 308$\pm15$ & 12$\pm3$ & 1.6$\pm0.2$ &SBb\cr
\+ $''$ (IRAM 30m) &&&   &  && --- & 380$\pm15$ & 16$\pm3$&2.5$\pm0.3$ &---\cr
\+IRAS 17527+6422&&& 26151 &26165$\pm20$&&  0.145 & 410$\pm20$ & 15$\pm5$ & 4.1$\pm0.4$& ?\cr

\vskip 2mm
\hrule
\vskip 4mm

\r $\dagger ~
W=W_{\rm rest} = c\Delta \nu_{\rm obs}/\nu_{\rm obs; center}=
c (1+z) \Delta \nu_{\rm obs}/(115.2712~{\rm GHz})$.

\endpage
\settabs 8\columns
\+ Table 3: Optical Imaging of three galaxies.$^\dagger$ \cr
\vskip 8mm
\hrule
\vskip 2mm

\+Galaxy&&Type&$V_{25.5}$&$R_{26.0}$&$I_{24.5}$&$A_{V_{25.5}}$
&$B_{V_{25.5}}$\cr
\+&&&(mag)&(mag)&(mag)&($''$)&($''$)\cr

\vskip 2mm
\hrule
\vskip 2mm

\+CGCG 1417.2+4759&&SBcd&15.34&14.82&14.39&32.2&23.1\cr
\+CGCG 1448.9+1654&&Sab&15.00&14.46&13.98&33.5&29.6\cr
\+CPG 60451&&Sc&15.97&15.31&14.72&35.2&13.7\cr

\vskip 2mm
\hrule
\v\v
\noi $\dagger$ The errors in the magnitudes are typically 0.03 mag.

\vskip 20mm

\settabs 8\columns
\+ Table 4. Data for the CO-Line TF Relation for measured galaxies.$^\dagger$\cr
\vskip 8mm
\hrule
\vskip 2mm

\+Galaxy&&$ V_{\rm T}$&$ R_{\rm T}$&$ I_{\rm T}$&$ V_C$&$i$&$ W_{i}$\cr
\+&&mag&mag&mag&\kms&deg&\kms \cr
\vskip 2mm
\hrule
\vskip 2mm
\+CGCG 1417.2+4759&&15.31&14.80&14.33&21502&45.3&433$\pm$27\cr
\+ $''~~''~~$ using  (IRAM 30-m)  && &    &     &     & &   534$\pm$33\cr
\+CGCG 1448.9+1654&&14.96&14.43&13.93&13483&28.7&560$\pm$54\cr
\+CPG 60451&&15.92&15.26&14.66&14840&70.1&489$\pm$18\cr

\vskip 2mm
\hrule
\v\v

\noi $\dagger$ The errors in the magnitudes are typically 0.1 mag,
and inclination errors are $\pm 3$ degrees.

\endpage

\settabs 8\columns
\def\p{$\pm$}
\def\kms{km s$^{-1}$}
\def\v{\vskip 2mm}
\noindent Table 5. Distance modulus, Distance,
and Hubble ratio for the detected galaxies. $^\dagger$
\vskip 8mm
\hrule
\vskip 2mm

\+Galaxy&&$ m-M$&&$D$ &&$H= <V_{\rm C}/r>$ \cr
\+      &&mag&& Mpc && \kms\ Mpc$^{-1}$ \cr
\vskip 2mm
\hrule
\vskip 2mm

\v

\+CGCG 1417.2+4759   (IRAM 30-m Telescope)\cr
\v
\+& ($V$-band)&     37.15$\pm$0.40  &&      270$\pm$50   &&   80$\pm$15 \cr
\+&  ($R_{\rm C}$-band)& 37.07$\pm$0.53  &&  260$\pm$62  &&   83$\pm$20\cr
\+&  ($I_{\rm C}$-band) &37.25$\pm$0.57   && 282$\pm$74 &&    76$\pm$20\cr
 \v

\+&(Avrg.)&     37.16$\pm$0.51& &  271$\pm$66 & &  80$\pm$19 \cr

\v
\+ CGCG 1448.0+1654 \cr
\v

\+& ($V$-band)& 36.93\p0.64 &&    243\p71      &&     55\p16\cr
\+& ($R_{\rm C}$-band)&  36.87\p0.84  &&  237\p92 &&  57\p16\cr
\+& ($I_{\rm C}$-band)& 37.03\p0.89&& 255\p103&&53\p21\cr
 \v

\+&(Avrg.)& 36.95\p0.79&&  245\p92&&   55\p21\cr
\v
\+ CPG 60451 \cr
\v
\+&  ($V$-band) & 37.53\p0.23&& 320\p34&& 46\p5\cr
\+&  ($R_{\rm C}$-band)  & 37.22\p0.31&&278\p39   && 53\p7\cr
\+&  ($I_{\rm C}$-band)  & 37.25\p0.33&&282\p43 &&  53\p8\cr
 \v

 \+&(Avrg.)&    37.33\p0.34 &&  293\p66 & &   51\p11\cr
\vskip 2mm
\hrule
\vskip 4mm
\r $\dagger$ The errors do not include systematic errors in photometric
corrections and internal scatter within the \tfr\ itself.

\vskip 20mm
\endpage

\noi{\bf Figure Captions}

\v\v\v

\r Fig. 1: CO line profiles for detected galaxies and  marginal
detection taken with the Nobeyama 45-m Telescope.
The CO line profile of galaxy CGCG 1417.2+4759 obtained with the IRAM 30-m
telescope is also presented.
All data are presented in $V_0$ vs $T_{\rm A}^*$  plane, where $V_0$
is radial velocity referred to the galaxy's optical $cz$.

\v\v 
\r Fig. 2: The same as Fig. 1, but non-detection or marginal for 
the farthest galaxies up to $cz=29000$ \kms\
in the present program obtained under good conditions.
Some show a sign of detection, while not conclusive as yet.

\v\v
\r Fig. 3: Optical images taken with the CFHT 3.6-m telescope
in $R_{\rm C}$ band of galaxies CGCG 1417.2+4759, 
CGCG 1448.9+1654, and CPG 60451 in contour map and  gray-scale representations.
Contours are drawn at an interval of 1 mag./square arcsec, starting
at 26.5 mag./square arcsec.

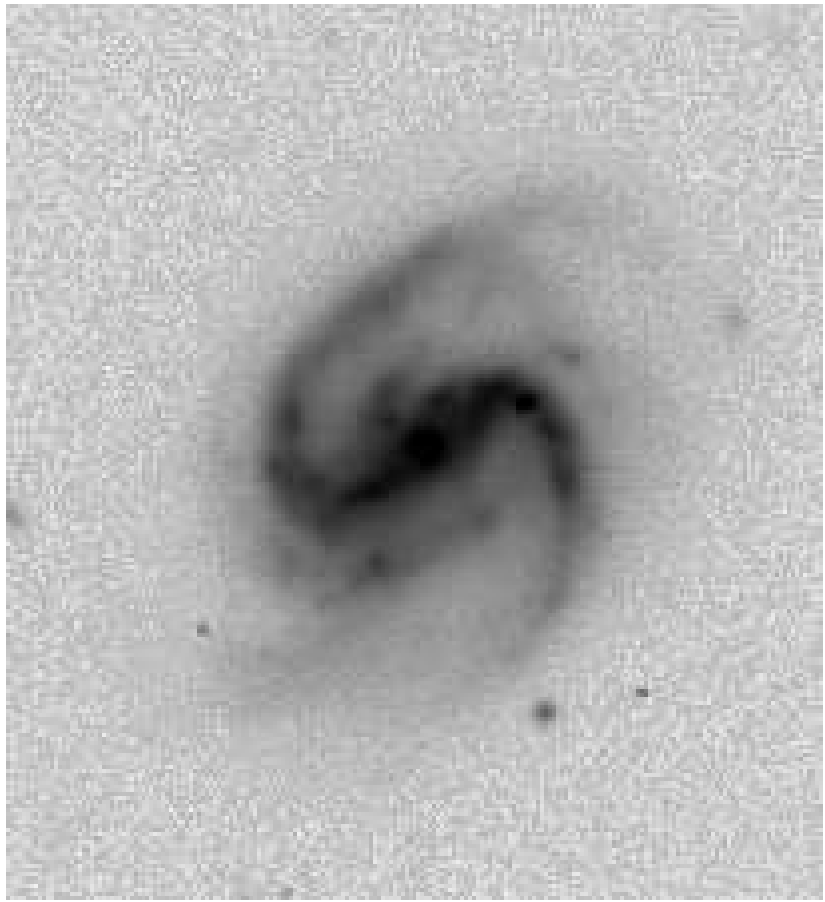
\bye